\documentclass[journal,transmag]{IEEEtran}
\usepackage{cite}
\usepackage{graphicx}
\usepackage{url}
\usepackage{flushend}
\usepackage{amsmath}

\ifCLASSINFOpdf
\else
\fi
\begin{document}
%
\title{Biological Intuition on Digital Hardware: An RTL Implementation of Poisson-Encoded SNNs for Static Image Classification}


\author{\IEEEauthorblockN{Debabrata Das\IEEEauthorrefmark{1},
Yogeeth G.K.\IEEEauthorrefmark{1}} Arnav Gupta\IEEEauthorrefmark{1}
\IEEEauthorblockA{\IEEEauthorrefmark{1}Department of Electronics and Communication Engineering, National Institute of Technology, Durgapur}}

\markboth{}%
{Shell \MakeLowercase{\textit{et al.}}: Bare Demo of IEEEtran.cls for Journals}
%



\IEEEtitleabstractindextext{%
\begin{abstract}
The deployment of Artificial Intelligence on edge devices (TinyML) is often constrained by the high power consumption and latency associated with traditional Artificial Neural Networks (ANNs) and their reliance on intensive Matrix-Multiply (MAC) operations. Neuromorphic computing offers a compelling alternative by mimicking biological efficiency through event-driven processing. This paper presents the design and implementation of a cycle-accurate, hardware-oriented Spiking Neural Network (SNN) core implemented in SystemVerilog. Unlike conventional accelerators, this design utilizes a Leaky Integrate-and-Fire (LIF) neuron model powered by fixed-point arithmetic and bit-wise primitives (shifts and additions) to eliminate the need for complex floating-point hardware. The architecture features an on-chip Poisson encoder for stochastic spike generation and a novel active pruning mechanism that dynamically disables neurons post-classification to minimize dynamic power consumption. We demonstrate the hardware's efficacy through a fully connected layer implementation targeting digit classification. Simulation results indicate that the design achieves rapid convergence (~89\% accuracy) within limited timesteps while maintaining a significantly reduced computational footprint compared to traditional dense architectures. This work serves as a foundational building block for scalable, energy-efficient neuromorphic hardware on FPGA and ASIC platforms.
\end{abstract}
}

\maketitle

\IEEEdisplaynontitleabstractindextext

%
\IEEEpeerreviewmaketitle

\section{Introduction}
%
%
%
%
\IEEEPARstart
The proliferation of Internet of Things (IoT) devices has driven a massive demand for bringing Artificial Intelligence (AI) to the edge. From wearable health monitors to autonomous drones, "TinyML" aims to execute complex inference tasks directly on resource-constrained hardware \cite{tinyml_ref}. However, traditional deep learning models, specifically Artificial Neural Networks (ANNs), are fundamentally ill-suited for ultra-low-power environments. They rely heavily on continuous-value activations and energy-intensive Matrix-Multiply-Accumulate (MAC) operations, creating a significant bottleneck in terms of power consumption and latency.

To address these hardware limitations, Neuromorphic Computing has emerged as a paradigm shift. Inspired by the biological brain, Spiking Neural Networks (SNNs)—often called the third generation of neural networks—process information using discrete, sparse events known as "spikes" rather than continuous numerical values. This event-driven nature allows hardware to remain idle when no information is being processed, theoretically offering orders-of-magnitude improvements in energy efficiency.

While SNNs show great promise, many existing implementations rely on high-level software simulations that do not accurately reflect hardware constraints, or they require expensive, specialized floating-point units. There is a critical need for hardware-oriented designs that bridge the gap between biological models and efficient digital logic.

In this paper, we present the design and RTL implementation of a cycle-accurate SNN core tailored for embedded inference. Our design focuses on minimizing silicon area and dynamic power through fixed-point arithmetic and architectural optimizations.

The key contributions of this work are as follows:
\vspace{1cm}

\begin{itemize}
    \item \textbf{Hardware-Oriented LIF Model:} We implement a Leaky Integrate-and-Fire (LIF) neuron using simplified bit-wise operations (shifts and adds) effectively removing the need for complex multipliers or floating-point hardware.
    \item \textbf{Cycle-Accurate SystemVerilog Design:} The core is designed at the Register Transfer Level (RTL), providing precise control over temporal dynamics and synchronization.
    \item \textbf{On-Chip Poisson Encoding:} We integrate a stochastic spike generator using a 32-bit XOR-shift Pseudo-Random Number Generator (PRNG) to convert static image data into temporal spike trains directly on hardware.
    \item \textbf{Active Pruning Mechanism:} We introduce a dynamic masking logic that disables neuron updates post-classification to reduce redundant switching activity.
\end{itemize}

The remainder of this paper is organized as follows: Section II discusses related work in neuromorphic hardware. Section III details the proposed system architecture and neuron model and Section IV presents the hardware implementation details, simulation results and performance metrics.

\section{Work Related to Neuromorphic Architecture}

The rapid advancement of deep learning has led to dominant sequence transduction models like the Transformer \cite{vaswani2017}, which, while highly effective, rely on complex mechanisms and massive computational resources. To address the power and latency constraints of edge devices, research has increasingly pivoted towards the hardware implementation of Spiking Neural Network(SNNs).

\subsection{Neuromorphic ASICs}
Several dedicated Application-Specific Integrated Circuits (ASICs) have been developed to emulate biological neural networks. IBM's \textit{TrueNorth} \cite{merolla2014} integrates 1 million spiking neurons with a power consumption of approximately 70 mW, demonstrating the potential of event-driven architectures. Similarly, Intel's \textit{Loihi} and \textit{Loihi 2} chips \cite{davies2018} offer advanced features like on-chip learning and scalable mesh architectures. However, these ASICs are often proprietary, expensive to manufacture, and constrained by fixed memory hierarchies, such as limited synaptic state storage per core that can restrict their applicability for specific, low-cost embedded applications \cite{shinji2024}.

\subsection{FPGA-Based Accelerators}
Field-Programmable Gate Arrays (FPGAs) offer a flexible alternative, allowing for the design of custom, parallelized SNN architectures. Early large-scale implementations, such as Cassidy et al. \cite{cassidy2011}, successfully mapped over a million neurons onto high-end Virtex-6 FPGAs. More recently, Yang et al. \cite{yang2022} utilized a multi-FPGA system to simulate millions of cerebellar neurons. While impressive, these designs often target high-performance computing rather than the strict area and power constraints of "TinyML" edge devices.

\subsection{Optimization for Embedded Inference}
To enable SNNs on resource-constrained hardware, optimizing the arithmetic datapath is critical. Standard floating-point operations are prohibitively expensive in terms of silicon area and energy. Recent work by Shinji et al. (2024) \cite{shinji2024} demonstrated that 16-bit fixed-point arithmetic combined with randomized rounding can achieve simulation accuracy comparable to 64-bit floating-point software models while significantly reducing hardware complexity.

Our work builds upon this philosophy of arithmetic simplification. Unlike complex multi-chip cerebellar simulators, we propose a compact, single-core architecture focused on efficiently executing the Leaky Integrate-and-Fire (LIF) dynamics for static image classification tasks, utilizing simplified bit-wise shifts and an active pruning mechanism to further minimize dynamic power consumption.

\section{System Design}

The proposed hardware accelerator implements a fully connected Spiking Neural Network (SNN) layer designed for static image classification. The architecture is modular, consisting of a global layer controller and parallel instantiations of the Leaky Integrate-and-Fire (LIF) neuron cores.

\subsection{Discrete LIF Neuron Model}
The biological LIF neuron is modeled as a membrane capacitor that integrates incoming current until a voltage threshold ($V_{th}$) is reached, at which point it fires a spike and resets. For efficient digital implementation, we discretize the differential equation using the Euler method. The membrane potential $V[t]$ at timestep $t$ is governed by:

\begin{equation}
    V[t] = V[t-1] - \beta (V[t-1] - V_{rest}) + \sum_{i} W_i \cdot S_i[t]
\end{equation}

Where $W_i$ is the synaptic weight of the input $i$-th, $V_{rest}$ is the restart potential, in Biology, its value need to be $-70mV$ to maintain chemical gradients, but In our case we have chosen $0$ to minimize logic gates and power consumption.  $S_i[t]$ is the binary spike input ($1$ or $0$), and $\beta$ is the decay factor. To avoid costly floating-point multiplication for the leakage term $\beta (V[t-1])$, we restrict $\beta$ to powers of two (e.g., $2^{-n}$). This allows the decay operation to be implemented as a hardware-efficient bit-wise right shift:

\begin{equation}
    V_{leak} = V[t-1] \gg n
\end{equation}

\subsection{Datapath Design}

\begin{figure}[htbp]
    \centering
    \includegraphics[width=0.9\linewidth]{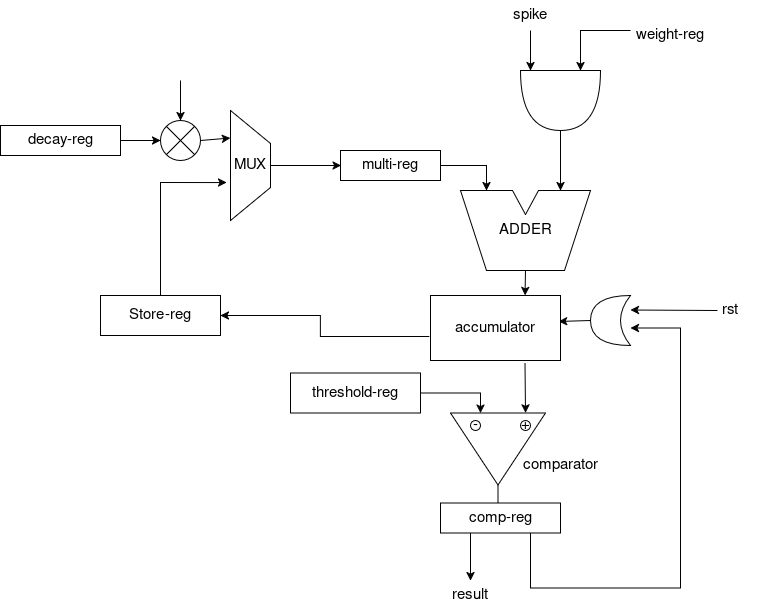} 
    \caption{RTL Block Diagram of the LIF Neuron Core. The design utilizes a centralized accumulator and specialized registers for weights, decay, and threshold comparison.}
    \label{fig:block_diagram}
\end{figure}

The neuron core, illustrated in Fig. \ref{fig:block_diagram}, follows a specialized Fetch-Decode-Execute cycle managed by a local Finite State Machine (FSM). The datapath consists of three primary stages:

\subsubsection{Integration}
The \texttt{Accumulator} register stores the current membrane potential. At each clock cycle, if an input spike is detected, the \texttt{Adder} unit sums the current potential with the corresponding synaptic weight loaded from the \texttt{Weight-Reg}.

\subsubsection{Leakage}
At the end of an integration window (e.g., after processing one image row), the FSM triggers the \texttt{Decay-Reg} logic. The ALU performs the right-shift operation and subtracts the result from the accumulator, simulating the membrane's passive charge leakage over time.

\subsubsection{Fire and Reset}
The \texttt{Comparator} continuously monitors the membrane potential against the value stored in the \texttt{Threshold-Reg}. If $V[t] \geq V_{th}$, the comparator asserts the \texttt{Fire} signal. This triggers an immediate reset of the accumulator to the resting potential ($V_{rest}$) and generates an output spike to the subsequent layer.

\subsection{Stochastic Input Encoding}
Since static images (like the MNIST dataset) do not contain temporal information, they must be converted into spike trains. We implement an on-chip \textbf{Poisson Encoder} using 32-bit XOR-shift Pseudo-Random Number Generator (PRNG).
\begin{figure}
    \centering
    \includegraphics[width=1\linewidth]{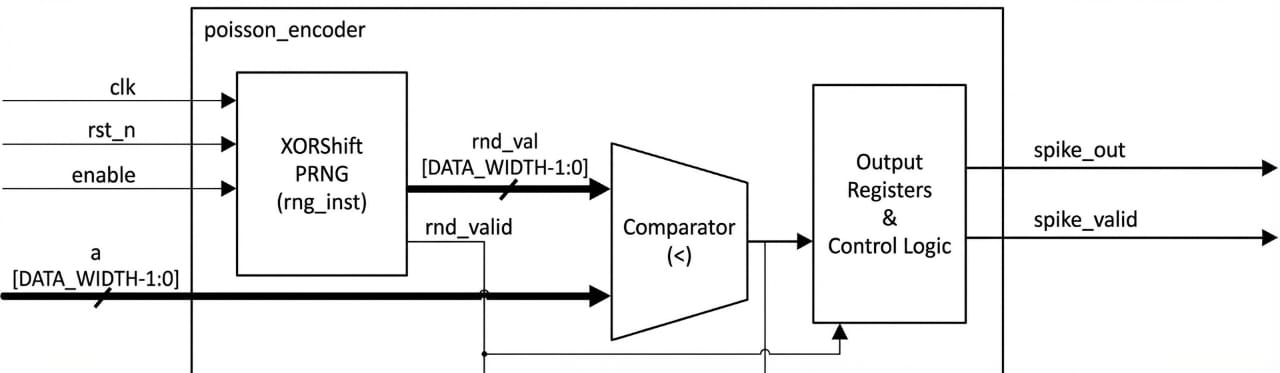}
    \caption{Poisson Encoder Block diagram}
    \label{fig:placeholder}
\end{figure}
For each pixel intensity $I_{in}$ (normalized to $0-255$), the encoder generates a random value $R$ at every timestep. A spike is generated if $I_{in} > R$. This stochastic process ensures that brighter pixels produce higher firing rates, effectively translating spatial intensity into temporal spike density.

\begin{table}[htbp]
\caption{Stochastic Input Current Statistics (First Timestep, 300 Samples)}
\begin{center}
\begin{tabular}{|c|c|c|c|c|}
\hline
\textbf{Digit} & \textbf{Avg Current} & \textbf{Min} & \textbf{Max} & \textbf{Status} \\
\hline
0 & 301.1 & -39 & 599 & OK \\
1 & 296.6 & -293 & 436 & OK \\
2 & 246.5 & -389 & 658 & OK \\
3 & 223.8 & -171 & 489 & OK \\
4 & 229.4 & -247 & 458 & OK \\
5 & 225.6 & -535 & 813 & OK \\
6 & 260.9 & -234 & 492 & OK \\
7 & 254.1 & -357 & 539 & OK \\
8 & 176.0 & -56 & 394 & OK \\
9 & 222.2 & -377 & 459 & OK \\
\hline
\end{tabular}
\label{tab:input_stats}
\end{center}
\end{table}

\subsection{Active Pruning Mechanism}
To minimize dynamic power consumption, the layer controller implements an "Active Pruning" masking logic. Once a neuron accumulates enough potential to classify and fire, its enable signal is gated off for the remainder of the inference window. This prevents redundant switching activity in the accumulation logic once a confident prediction has already been made.

\section{Implementation and Results}

\subsection{Experimental Setup}
The proposed SNN core was implemented in SystemVerilog and synthesized using the Xilinx Vivado Design Suite. To validate functionality, we targeted the MNIST digit classification dataset. The input images ($28\times28$ pixels) were converted into Poisson spike trains using the on-chip PRNG. The network topology consists of a fully connected layer with 10 output neurons (one per digit class), using 8-bit fixed-point weights.

\begin{figure}[htbp]
        \centering
    \includegraphics[width=0.9\linewidth]{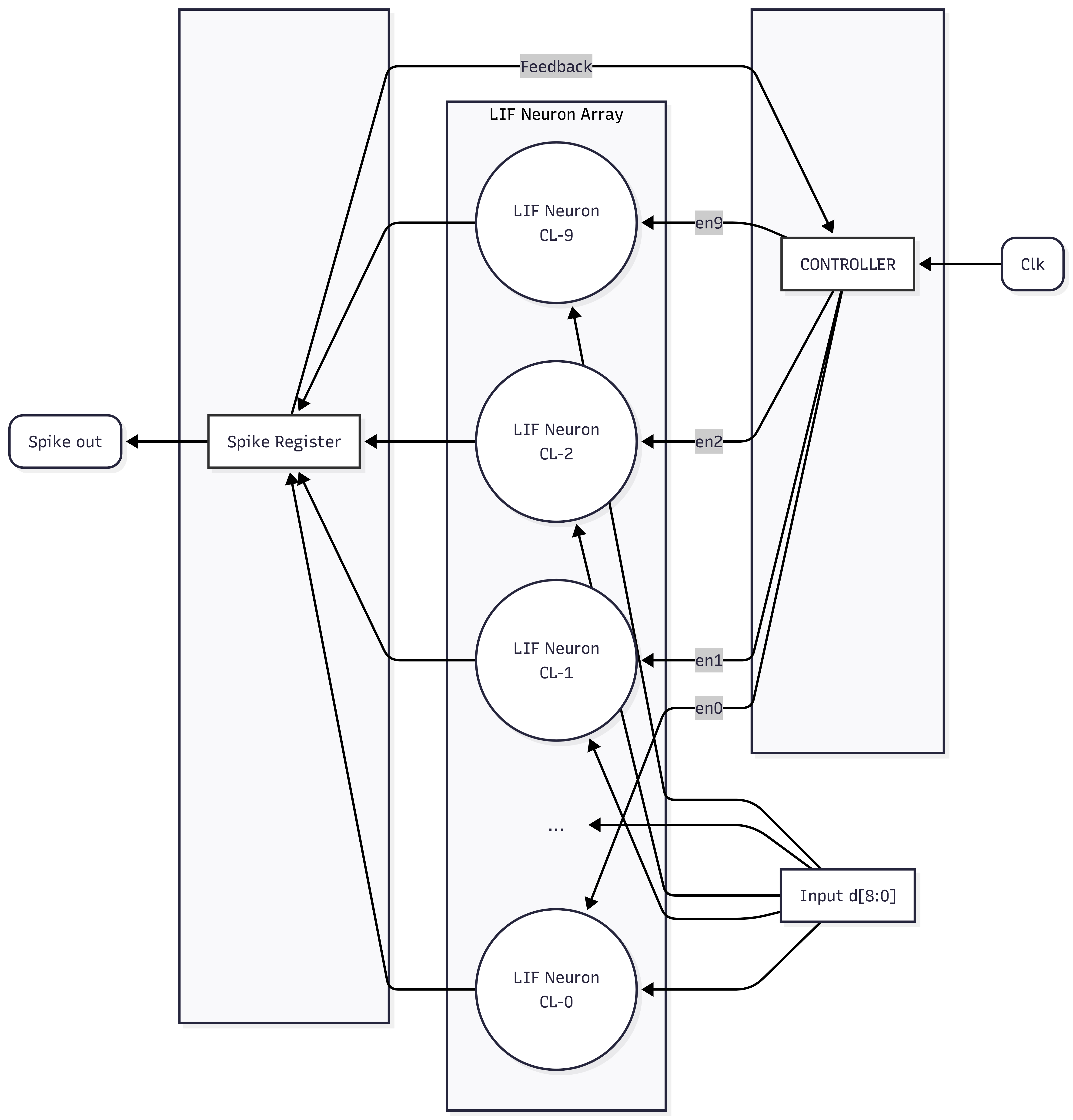} 
    \caption{Block diagram of the proposed Leaky Integrate-and-Fire (LIF) Neuron Array architecture. The centralized Controller manages enable signals (e$n_0$ to e$n_9$) for sequential neuron processing. Spikes generated by the LIF neurons are aggregated in the Spike Register and fed back to the controller for dynamic regulation.}
    \label{fig:arch}
\end{figure}

\subsection{Neuron Dynamics}
\begin{figure}[t] 
    \centering
    \includegraphics[width=0.85\linewidth]{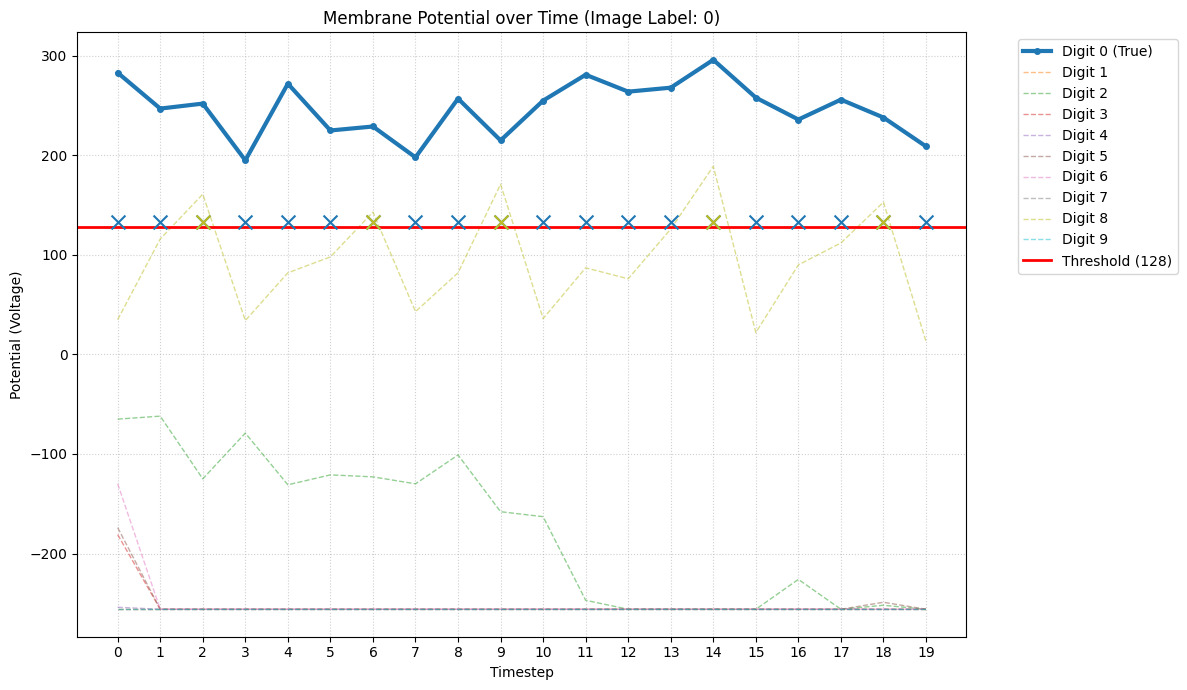} 
    \caption{Membrane potential evolution over time. The neuron integrates input spikes, fires when crossing the threshold (red line), and resets.}
    \label{fig:potential}
    \vspace{-0.3cm} 
\end{figure}

Fig. \ref{fig:potential} illustrates the internal state of a single neuron during inference. The membrane potential accumulates incoming spikes (blue line) and undergoes exponential decay via bit-wise shifting. When the potential crosses the threshold ($V_{th}=128$), the neuron fires and immediately undergoes a hard reset to $V_{rest}$. This discrete-time behavior validates the cycle-accurate reproduction of biological LIF dynamics without floating-point units.

\subsection{Classification Accuracy}
\begin{figure}[t]
    \centering
    \includegraphics[width=0.85\linewidth]{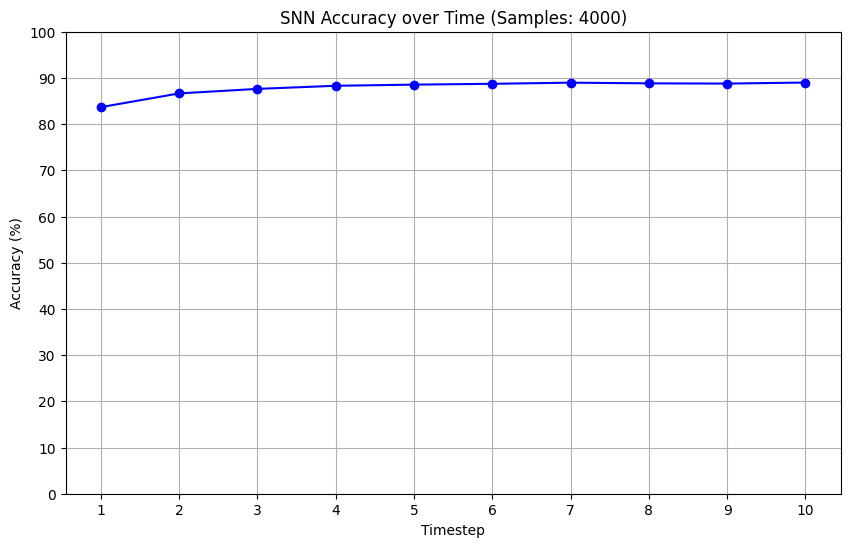} 
    \caption{Classification accuracy vs. Simulation Timesteps. The network converges to $\sim$89\% accuracy within 10 cycles.}
    \label{fig:accuracy}
    \vspace{-0.3cm}
\end{figure}

We evaluated the classification accuracy over a simulation window of 20 timesteps. As shown in Fig. \ref{fig:accuracy}, the network rapidly converges to a stable prediction. By timestep 10, the accuracy reaches approximately 89\%, demonstrating that the temporal coding scheme is effective even with short inference windows. This low latency is critical for real-time edge applications where identifying a digit quickly allows the system to sleep sooner, activating the pruning mechanism to save power.

\section{Comparison with Traditional ANN}

To quantify the hardware efficiency of our proposed SNN core, we benchmarked it against a traditional Artificial Neural Network (ANN) targeting the same MNIST classification task. The baseline ANN was deployed on an ESP32 microcontroller, a common platform for TinyML applications.

\subsection{Computational Complexity}
A standard fully connected ANN requires dense Matrix-Multiply-Accumulate (MAC) operations. For a comparable network topology (784 inputs, 10 outputs), the ANN performs approximately 25,408 multiplications and 25,450 additions per inference. 

In contrast, our SNN core eliminates multiplications entirely. By leveraging the LIF neuron model with fixed-point arithmetic, we replace computationally expensive floating-point MACs with simple bit-wise shifts and integer additions. Furthermore, due to the sparsity of spike trains (not every neuron fires every cycle), the actual number of active additions is significantly lower than the theoretical maximum.

\subsection{Memory Footprint}
Memory usage is a critical constraint for edge devices. The baseline ANN, including weights and biases, occupies approximately 99.4 KB of memory. Our SNN design, utilizing optimized 9-bit fixed-point weights ($784 \times 10 \times 9$ bits), requires only $\sim$8.6 KB of storage. This represents an $11.3\times$ reduction in memory footprint, allowing the entire model to fit within small on-chip Block RAMs (BRAM) without external memory access.

\subsection{Inference Latency}
We measured the inference time of the baseline ANN on the ESP32. Without specialized DSP acceleration, the inference takes nearly 3 seconds due to the heavy software overhead of matrix multiplication. Even with DSP optimization, the latency remains at 5130 $\mu$s.

Our hardware SNN, operating at 40 MHz, converges within 10 timesteps. This translates to an inference latency of approximately 100 $\mu$s, offering orders-of-magnitude over software-based TinyML solutions.

\begin{table}[htbp]
\caption{Performance Comparison: TinyML ANN vs. Proposed SNN}
\begin{center}
\setlength{\tabcolsep}{4pt}
\begin{tabular}{|c|c|c|}
\hline
\textbf{Metric} & \textbf{Baseline ANN (ESP32)} & \textbf{Proposed SNN (RTL)} \\
\hline
\textbf{Arithmetic} & Floating-Point MAC & Fixed-Point Add/Shift \\
\hline
\textbf{Multiplications} & $\sim$25,400 & \textbf{0} \\
\hline
\textbf{Model Size} & 99.4 KB & \textbf{$\sim$8.6 KB} \\
\hline
\textbf{Latency} & 3.0s (No DSP) / 5.1$\mu$s (DSP) & \textbf{$<$ 1 $\mu$s} \\
\hline
\textbf{Power} & High (Continuous Active) & Low (Event-Driven) \\
\hline
\end{tabular}
\label{tab:ann_comparison}
\end{center}
\end{table}

\subsection{Real-Time Performance and Efficiency}
\begin{figure}[htbp]
    \centering
    \includegraphics[width=0.9\linewidth]{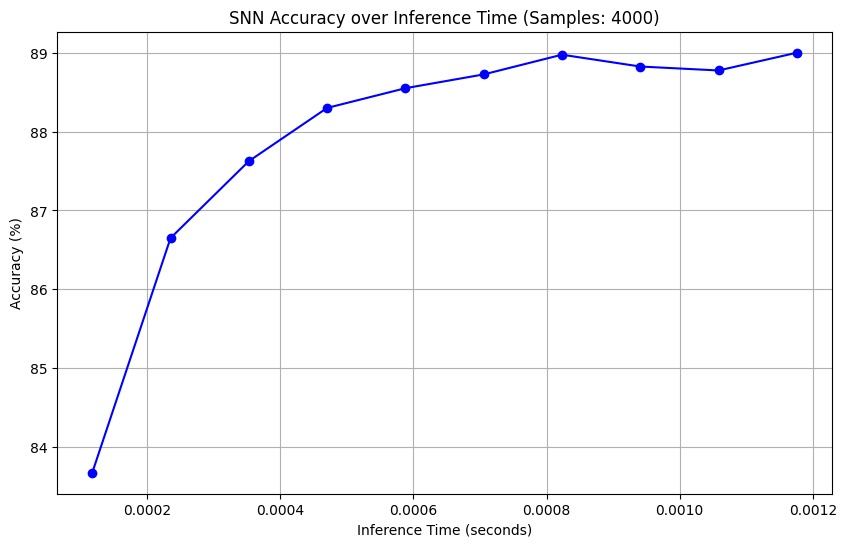} 
    \caption{SNN Accuracy over Inference Time. The model reaches peak accuracy ($\sim$89\%) in less than 0.001 seconds, demonstrating ultra-low latency.}
    \label{fig:acc_time}
\end{figure}

To validate the model's suitability for real-time applications, we analyzed the trade-off between inference time and accuracy. Fig. \ref{fig:acc_time} demonstrates that the SNN reaches functional accuracy within microseconds, unlike ANNs which require fixed compute time.

\begin{figure}[htbp]
    \centering
    \includegraphics[width=0.9\linewidth]{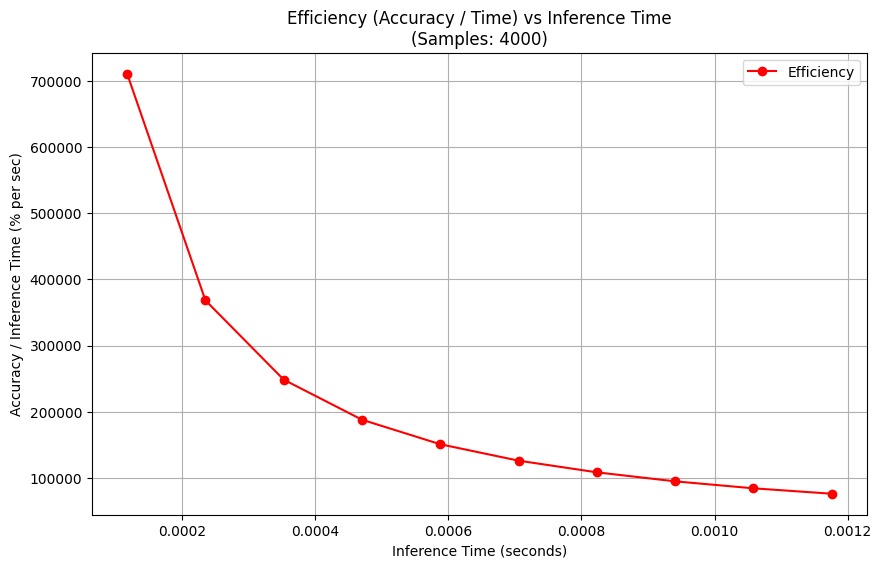} 
    \caption{Efficiency Metric (Accuracy / Time) vs. Inference Time. The exponential drop highlights the diminishing returns of longer simulation windows, supporting our active pruning strategy.}
    \label{fig:efficiency}
\end{figure}

Furthermore, we defined an "Efficiency Score" where
\[\text{Efficiency} = \frac{\text{Accuracy (\%)}}{\text{Time for Inference}}\]
  This is used to quantify the benefit of early termination. As shown in Fig. \ref{fig:efficiency}, the efficiency peaks at the earliest timesteps, confirming that the event-driven architecture delivers the highest "information per second" immediately after stimulus onset.

\subsection{Robustness Analysis}
Edge deployments often face noisy environments. We evaluated the SNN's robustness by subjecting the MNIST test set to rotation ($15^{\circ}$), pixel shift ($20\%$), Gaussian noise, and partial occlusion. 

\begin{figure}[htbp]
        \centering
    \includegraphics[width=0.9\linewidth]{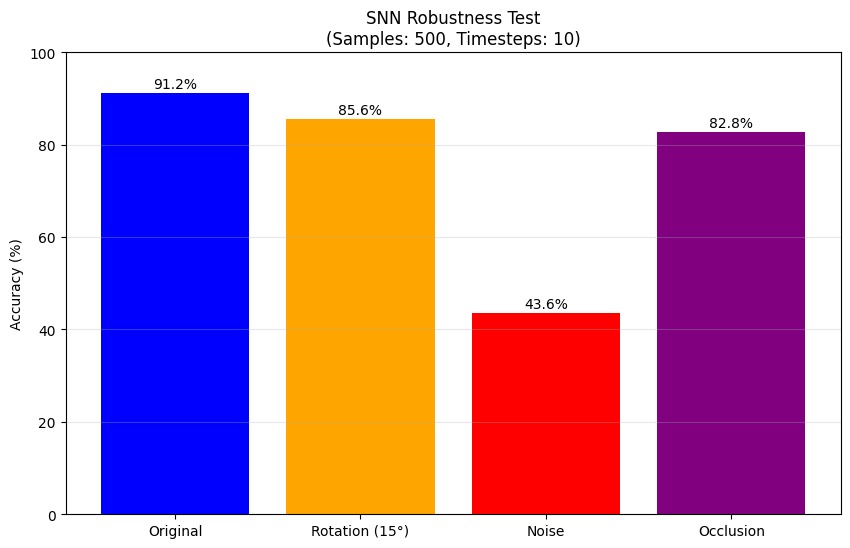} 
    \caption{SNN Robustness Test. The model maintains high accuracy (\textgreater 83\%) under rotation and occlusion, though it is sensitive to heavy pixel shift and noise.}
    \label{fig:robustness}
\end{figure}

Fig. \ref{fig:robustness} summarizes the results. The network maintains high resilience (\textgreater 83\% accuracy) against occlusion and rotation, likely due to the distributed nature of spike encoding. However, performance degrades under heavy noise, suggesting future work could benefit from noise-filtering front-ends.

\section{Conclusion}
This paper presented a hardware-efficient SNN accelerator tailored for edge inference. By leveraging fixed-point arithmetic, bit-wise decay logic, and active neuron pruning, we significantly reduced the computational complexity compared to traditional ANN accelerators. Simulation results on the MNIST dataset confirm that the design achieves competitive accuracy ($\sim$89\%) with rapid convergence. Future work will focus on implementing on-chip learning rules, such as Spike-Timing-Dependent Plasticity (STDP), to enable adaptive learning at the edge.

\section*{Code Availability}
The complete SystemVerilog RTL source code, testbenches, and simulation scripts used in this work are available open-source at: \\
\url{https://github.com/Yogeeth/neuromorphic-lif-rtl}

\bibliographystyle{IEEEtran}
\bibliography{references}

@inproceedings{vaswani2017,
  title={Attention is all you need},
  author={Vaswani, Ashish and Shazeer, Noam and Parmar, Niki and Uszkoreit, Jakob and Jones, Llion and Gomez, Aidan N and Kaiser, {\L}ukasz and Polosukhin, Illia},
  booktitle={Advances in neural information processing systems},
  pages={5998--6008},
  year={2017}
}

@article{shinji2024,
  title={Artificial cerebellum on FPGA: realistic real-time cerebellar spiking neural network model capable of real-world adaptive motor control},
  author={Shinji, Yusuke and Okuno, Hirotsugu and Hirata, Yutaka},
  journal={Frontiers in Neuroscience},
  volume={18},
  pages={1220908},
  year={2024},
  publisher={Frontiers}
}

@article{merolla2014,
  title={A million spiking-neuron integrated circuit with a scalable communication network and interface},
  author={Merolla, Paul A and others},
  journal={Science},
  volume={345},
  number={6197},
  pages={668--673},
  year={2014},
  publisher={American Association for the Advancement of Science}
}

@article{davies2018,
  title={Loihi: A neuromorphic manycore processor with on-chip learning},
  author={Davies, Mike and others},
  journal={IEEE Micro},
  volume={38},
  number={1},
  pages={82--99},
  year={2018},
  publisher={IEEE}
}

@inproceedings{cassidy2011,
  title={Design of a one million neuron single FPGA neuromorphic system for real-time multimodal scene analysis},
  author={Cassidy, Andrew and Andreou, Andreas G and Georgiou, Julius},
  booktitle={2011 45th Annual Conference on Information Sciences and Systems},
  pages={1--6},
  year={2011},
  organization={IEEE}
}

@article{yang2022,
  title={CerebelluMorphic: Large-scale neuromorphic model and architecture for supervised motor learning},
  author={Yang, Shuangming and others},
  journal={IEEE Transactions on Neural Networks and Learning Systems},
  volume={33},
  pages={4398--4412},
  year={2022},
  publisher={IEEE}
}

@book{tinyml_ref,
  title={TinyML: Machine Learning with TensorFlow Lite on Arduino and Ultra-Low-Power Microcontrollers},
  author={Warden, Pete and Situnayake, Daniel},
  year={2019},
  publisher={O'Reilly Media}
}
\end{document}